\documentclass[aps,prl,twocolumn,showpacs,superscriptaddress]{revtex4}
\usepackage{graphicx,color,epsfig}
\usepackage{amssymb,amsmath}

\begin{document}
\title{Localized coherence in two interacting populations of social agents}
\author{J. C. Gonz\'alez-Avella}
\affiliation{Instituto de F\'isica, Universidade Federal do Rio Grande do Sul, 91501-970 Porto Alegre, Brazil}
\affiliation{Instituto Nacional de Ci\^encia e Tecnolog\'ia de Sistemas Complexos, INCT-SC, 91501-970 Porto Alegre, Brazil}
\author{M. G. Cosenza}
\affiliation{Centro de F\'isica Fundamental, Universidad de los Andes, M\'erida,  M\'erida 5251, Venezuela.}
\author{M. San Miguel}
\affiliation{IFISC Instituto de F\'isica Interdisciplinar y
Sistemas Complejos (CSIC-UIB), E-07122 Palma de Mallorca, Spain}

\begin{abstract}
We investigate the emergence of localized coherent behavior in  systems consisting of 
two populations  of social agents possessing a condition for non-interacting states, mutually coupled through global interaction fields.  
We employ two examples of such dynamics: (i) Axelrod's model for social influence, and (ii) a discrete version of a bounded confidence
model for opinion formation. 
In each case, the global interaction fields correspond
to the statistical mode of the states of the agents in each population.
In both systems we find localized coherent states for some values of parameters,
consisting of one population in a homogeneous state and the other in a disordered state. This situation can be considered 
as a social analogue to a chimera state
arising in two interacting populations of oscillators. 
In addition, other asymptotic collective behaviors appear in both systems depending on parameter values: 
a common homogeneous state, where both populations reach the same state;
different homogeneous states, where both population reach homogeneous states different from each other;
 and a disordered state, where both populations reach inhomogeneous states.
\end{abstract}
\maketitle

The study of the collective behaviors in systems consisting of two interacting populations of dynamical elements is a topic of much interest in various sciences. These systems are characterized by the presence of nonlocal interactions between elements in different populations. Examples of such systems arise in the coexistence of biological species \cite{Montro,Gomatan,Dubey}, the competition of  two languages in space \cite{Patri}, 
and in the dynamics of two networks of coupled oscillators \cite{Kurths,Barreto,Abrams}. 

Recently, a remarkable phenomena called chimera \cite{Kuramoto,Strogatz} has been found 
in systems consisting of two populations of oscillators 
subject to reciprocal interactions \cite{Abrams,Laing,Scholl}. In a chimera state, one population  
exhibits a coherent or synchronized behavior while the other is incoherent or desynchronized. 
The recent experimental discovery of such chimera states has fundamental implications as it shows that localized coherence and structured patterns can emerge from otherwise structureless systems \cite{Showalter,Roy}. 
As noted in Ref.~\cite{Abrams}, analogous symmetry breaking is observed in dolphins and other animals
that have evolved to sleep with only half of their brain at a time: neurons exhibit
synchronized activity in the sleeping hemisphere and desynchronized activity in the
hemisphere that is awake \cite{Lima}. 

In this paper we investigate the 
emergence of localized coherence in 
 systems consisting of two populations of social agents  
coupled through reciprocal global interactions.    
In a first system, 
we employ, as interaction dynamics, Axelrod's \cite{Axel} rules for
the dissemination of culture among agents in a society, 
a model that has attracted much attention from physicists \cite{CMV,Klemm2,JC1,JC,Kuperman,Candia,Galla,Gracia,Zhang}. 
In the second system that we consider, we introduce a discrete version of the bounded confidence model proposed by Deffuant et al. \cite{Deff},
where agents can influence each other's opinion provided that opinions are already sufficiently close enough. 
In both models of interaction dynamics, the agent-agent interaction rule is such that no interaction exists
for some relative values characterizing the states of the agents that compose the system. This
type of interaction is common in social and biological systems where there is often some bound or threshold
for the occurrence of interaction between agents, such as a similarity condition for the state
variable \cite{Deff,Mikhailov,Amblard,Krause,Zanette}.
The global interactions act as fields  \cite{JC} that can be
interpreted as mass media messages originated in each population. 
Thus, our system can serve as a model for cross-cultural interactions between two social groups, each with its own internal dynamics, but getting  information 
about each other 
through their reciprocal mass media influences \cite{Plos1}.  In particular, the study of cross-cultural experiences 
through mass-mediated contacts is a relevant issue in the Social Sciences \cite{Bryant,Yaple,Chinos}. 

We show that, in both models and under some circumstances, one population  reaches a homogeneous state while 
a disordered state appears on the other. This configuration is similar to a chimera state
arising in two populations  of oscillators subject to  global interactions.  

As model I, we consider a system of $N$ agents divided into two populations: $\alpha$ and $\beta$,  with sizes $N_\alpha$ and $N_\beta$, such that $N=N_\alpha+N_\beta$.
Each population consists of a fully connected network, i. e., every agent can interact with any other within a population.
We use the notation $[z]$ to indicate ``or $z$''.
The state of agent $i \in \alpha [\beta]$ is given by an $F$-component vector $x_{\alpha[\beta]}^f(i)$,
$(f = 1, 2, \ldots,F)$, where each component can take any of $q$ different values 
$x_{\alpha[\beta]}^f(i) \in \{0, 1, . . . , q-1\}$. 
Here we define the normalized parameter $Q \equiv 1-(1-1/q)^F$ to express the decreasing number of initial options per component, 
such that $Q=0$ for $q\to\infty$ (many options), and $Q=1$ for $q=1$ (one option). 

We denote by $M_\alpha=(M_\alpha^1,\ldots,M_\alpha^f,\ldots,M_\alpha^F)$ and $M_\beta=(M_\beta^1,\ldots,M_\beta^f,\ldots,M_\beta^F)$ 
the global fields  defined as the statistical modes of the states in the populations $\alpha$ and  $\beta$, respectively, at a given time. 
Thus, the component $M_{\alpha[\beta]}^f$ 
is assigned the most abundant value exhibited by the $f$th component of all the state vectors $x_{\alpha[\beta]}^f(i)$ in the population $\alpha[\beta]$. If the maximally abundant value is not unique, one of the possibilities
is chosen at random with equal probability. 
In the context of social dynamics, these global fields correspond
to cultural ``trends'' associated to each population.
Each agent in population $\alpha$ is subject to the influence of the global field $M_\beta$, and  each agent in population $\beta$ is subject to the influence of the global field $M_\alpha$. Then, the global fields can be interpreted as reciprocal mass media messages originated in one population and being transmitted to the other.  

The states $x_{\alpha[\beta]}^f(i)$ are initially assigned at random with a uniform distribution in each population. 
At any given time, a randomly selected agent in population $\alpha[\beta]$ 
can interact either with the global field  $M_{\beta[\alpha]}$ or with any other agent belonging to 
$\alpha[\beta]$, 
in each case according to the dynamics of Axelrod's cultural model.
The dynamics of the system is defined by the following iterative algorithm:
\begin{enumerate}
\item Select at random an agent $i\in \alpha$ and a agent $j \in \beta$.
\item Select the source of interaction:
with probability $B$, agent $i \in \alpha$ interacts with field $M_\beta$ and agent $j \in \beta$ interacts with field $M_\alpha$, while
with probability $1-B$,  $i$ interacts with $k \in \alpha$ selected at random and $j$ interacts with $l \in \beta$ also selected at random. 
\item Calculate the overlap, i. e., the number of shared components, between the state of agent $i \in \alpha$ and the state of its source of interaction, defined by $d_\alpha(i,y)= \sum_{f=1}^F \delta_{x_\alpha^f(i),y^f}$, where $y^f=M_\beta^f$ if the source is the field $M_\beta$, or $y^f=x_\alpha^f(k)$ if the source is agent $k \in \alpha$. Similarly, calculate the overlap $d_\beta(j,y)= \sum_{f=1}^F \delta_{x_\beta^f(j),y^f}$, where $y^f=M_\alpha^f$ if the source is the field $M_\alpha$, or $y^f=x_\beta^f(l)$ if the source is agent $l\in \beta$. Here we employ the delta Kronecker function, $\delta_{x,y}=1$, if $x=y$;  $\delta_{x,y}=0$, if $x\neq y$.
\item If $0 < d_\alpha(i,y) <F$, with probability $\frac{d_\alpha(i,y)}{F}$ choose $g$ such that $x_\alpha^g(i)\neq y^g$ and set 
$x_\alpha^g(i)=y^g$; if $d_\alpha(i,y)=0$ or $d_\alpha(i,y)=F$, the state $x_\alpha^f(i)$ does not change. 
If $0 < d_\beta(j,y) <F$, with probability $\frac{d_\beta(j,y)}{F}$ choose $h$ such that $x_\beta^h(j)\neq y^h$ 
and set $x_\beta^h(j)=y^h$; if $d_\beta(j,y)=0$  or  $d_\beta(j,y)=F$, the state $x_\beta^f(j)$ does not change.
\item If the source of interaction is $M_{\beta[\alpha]}$, update the fields $M_\alpha$ and $M_\beta$.
\end{enumerate}

In step $(2)$, the parameter $B \in [0, 1]$ describes the
probability for the agent-field interactions and represents the strength of the fields $M_\alpha$ and $M_\beta$. 
Steps $(3)$ and $(4)$ describe the interaction rules from Axelrod's model for social influence. 
Step $(5)$ characterizes the time scale for the updating of the global fields.
The non-instantaneous updating of the global fields expresses the delay with which 
a population  acquires
knowledge about the other through
the only available communication channel between them,
as described in many societies experiencing cross-cultural interactions through mass media \cite{Bryant}. 

In the asymptotic state, both populations $\alpha$ and $\beta$ form domains of different sizes.
A domain is a set of connected agents that share the same state.
A homogeneous state in population $\alpha[\beta]$ is characterized by $d_\alpha(i,j)=F$, $\forall i, j \in \alpha[\beta]$. There are $q^F$ 
equivalent configurations for this state.  The coexistence of several domains in a population corresponds to an inhomogeneous or disordered state.

For $B = 0$,  we have two uncoupled and independent populations. It is known \cite{CMV,Klemm2} that   
a single system subject to Axelrod's dynamics asymptotically 
reaches a homogeneous phase
for values $q < q_c$, and a disordered phase
for $q > q_c$, where $q_c$ is a critical point. For fully connected networks, the value $q_c$ depends on the system size \cite{Fede}. 
In terms of the parameter $Q$, the disordered phase occurs for $Q < Q_c = 1-(1-1/q_c)^F$
and the homogeneous phase takes place for $Q > Q_c$.

\begin{figure}[h]
\begin{center}
\includegraphics[width=0.7\linewidth,angle=0]{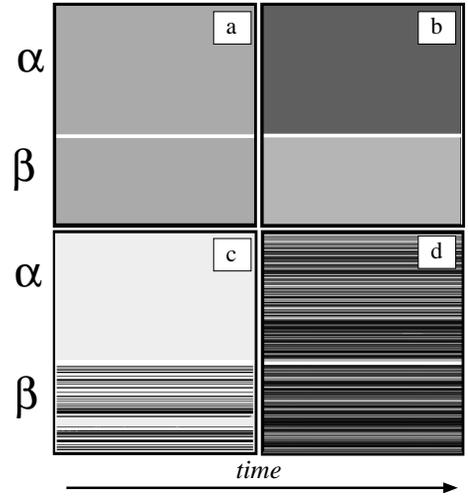}
\end{center}
\caption{Model I. Asymptotic states (vertical axis) of the agents in interacting populations $\alpha$ (upper part) and $\beta$  (lower part) vs. time (horizontal axis) represented in gray colors, for different values of parameters, with fixed $F=10$. 
Each vector state variable of an agent is represented by a different shade of gray. Population sizes are $N_\alpha=0.6N$, $N_\beta=0.4N$, with $N=800$.
(a) $B=0.001, \,Q=0.118$ (common homogeneous state). (b) $B=0.001, \, Q=0.095$ (different homogeneous states).
(c) $B=0.05, \, Q=0.12$ (localized coherent state). (d) $B=0.25, \,  Q=0.004$ (disordered state).}
\label{F1}
\end{figure}

As the intensity of the global fields $B$ is increased, the system exhibits diverse asymptotic behaviors for different
values of the parameter $Q$.  Figure~\ref{F1} displays the asymptotic spatiotemporal patterns corresponding to the main
behaviors observed:
(a) a common homogeneous state, where both populations reach the same state, $M_\alpha=M_\beta$;
(b) different homogeneous states, where both population reach homogenous states different from each other,
 $M_\alpha \neq M_\beta$;
(c) localized coherent state, where a homogeneous state occurs in only one population while the other is inhomogeneous; 
and (d) a disordered state, where both populations reach inhomogeneous states for values $Q < Q_c$.

The collective behaviors of the system can be characterized by employing the following statistical quantities:
(i) the average normalized size (divided by $N_{\alpha[\beta]}$) of the largest domain in $\alpha[\beta]$, denoted by $S_{\alpha[\beta]}$; 
(ii) the probability that the largest domain in $\alpha[\beta]$ has a state equal to $M_{\beta[\alpha]}$, designed by $P_{\beta[\alpha]}(M_{\alpha[\beta]})$;
(iii) the probability $\phi$ of finding a localized coherent state in the system (either population coherent, the other
disordered).

\begin{figure}[h]
\begin{center}
\includegraphics[width=0.9\linewidth,angle=0]{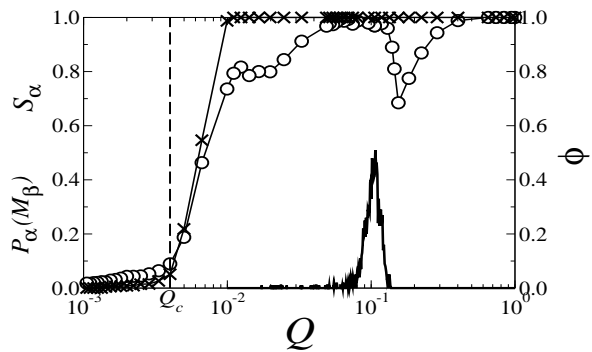}
\end{center}
\caption{Model I. Quantities $S_\alpha$,  $P_\alpha(M_\beta)$, and the probability $\phi$ of finding a localized ordered state in the system, as functions of the parameter $Q$ with $F=10$, for fixed $B=0.05$.  Each data point is the result of averaging over $100$ realizations of initial conditions. 
System size is $N=800$ with partition $N_\alpha=0.6N$. Left vertical axis: $S_\alpha$ (open circles); $P_\alpha(M_\beta)$ (crosses). Right vertical axis: probability $\phi$ (continuous thick line). Disordered states occur for values $Q < Q_c=0.004$.}
\label{F2}
\end{figure}

Figure~\ref{F2} shows these quantities as functions of the parameter $Q$, for a fixed value $B=0.05$.
The qualitative behavior of the system is similar for other values of $B$ and also for different sizes of the
partitions of the two populations.  
The probability $P_\alpha(M_\beta)=1$ for values $Q > Q_c=0.004$,
indicating that the state of the largest domain in $\alpha$ is always equal to that imposed by the field $M_\beta$. 
For values  $Q$ close to $1$, each population reaches a homogeneous state with $S_{\alpha[\beta]}=1$,  
This means that, for this range of $Q$, 
the global field $M_\beta$ imposes its state on population $\alpha$ 
and, correspondingly, the field  $M_\alpha$ imposes its state on population $\beta$. Consequently, both populations 
reach the same homogeneous state with  $M_\alpha=M_\beta$. This asymptotic state 
is shown in Fig.~\ref{F1}(a). 
However, for very small values of $B$, the spontaneous coherence arising in population $\alpha$ for parameter
 values $Q > Q_c$ due to the agent-agent interactions competes with the order being imposed
 by the applied global field $M_\beta$. For some realizations of initial conditions,
the homogenous state in population $\alpha[\beta]$ does not always 
coincides with the state of the applied global field $M_{\beta[\alpha]}$. In that case, 
populations $\alpha$ and $\beta$ may reach homogeneous states different from each other, where
$M_\beta \neq M_\alpha$. This state is displayed in Fig.~\ref{F1}(b).
For values $Q < Q_c$, $\forall B$, both populations reach disordered states, characterized by 
$S_\alpha \simeq  S_\beta \simeq  0$. The corresponding pattern is exhibited in Fig.~\ref{F1}(d).

Note that $S_\alpha<1$ for some ranges of values of $Q$, indicating that for those values the largest domain in population $\alpha$ 
does not entirely occupy that population. This corresponds to a state of partial coherence for
both populations. 

As shown in Fig.~\ref{F1}(c), localized coherent states are  configurations where a homogeneous state can arise in only one population, 
while the other remains inhomogeneous. 
In contrast to the other homogeneous states that can be characterized by statistical quantities calculated in just one population, a localized coherent state is defined
by considering both populations simultaneously, i.e., it requires the observation of the entire system.
A localized coherent state is characterized by
$S_{\alpha[\beta]}=1$ and $S_{\beta[\alpha]}=u<1$, where $u$ is some threshold value.
Figure~\ref{F2} shows the probability $\phi$ of finding a localized coherent state in the system as a function of $q$, employing the criterion $u \leq 0.6$. There are ranges of the parameter $Q$ where localized coherent states can emerge;  
the probability $\phi$ 
is maximum immediately before the value of $Q$ corresponding to a local minimum of $S_\alpha$. 
Note that the region of the parameter $Q$ where localized coherent states appear in the system lie between a
common homogeneous state and a partially coherent state. The configuration of localized coherent states shares features of both of these states.\\

\begin{figure}[h]
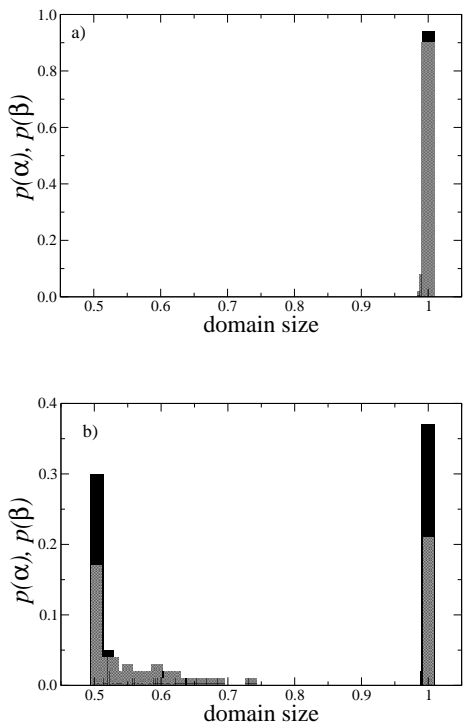

\begin{center}
\includegraphics[width=0.7\linewidth,angle=0]{F3_A.eps}
\end{center}
\vspace{1mm}
\begin{center}
\includegraphics[width=0.7\linewidth,angle=0]{F3_B.eps}
\end{center}
\caption{Model I. Probability distributions  $p(\alpha)$ (black bars) and $p(\beta)$ (gray bars) of normalized domain sizes for populations $\alpha$ and $\beta$, calculated over 
$100$ realizations of initial conditions,
with fixed $B=0.05$ and for different values of $Q$. (a) $Q=0.65$ (common homogenous state); (b) $Q=0.105$ (localized coherent states).}
\label{F3}
\end{figure}

Figure~\ref{F3} shows the probability distributions $p(\alpha)$ and $p(\beta)$ of the normalized domain sizes for populations $\alpha$ and $\beta$, respectively, calculated over $100$ realizations of initial conditions, for different values of $Q$, and with fixed $B=0.05$ corresponding to Fig.~2. 
Figure~\ref{F3}(a) exhibits the probabilities $p(\alpha)$ and $p(\beta)$ with $Q=0.65$, corresponding to a common homogeneous state characterized by the presence of
one large domain in each population whose size is of the order of that population size $S_\alpha[S_\beta] \sim 1$.  
Figure~\ref{F3}(b) shows  $p(\alpha)$ and $p(\beta)$ for $Q=0.105$, corresponding to the emergence of localized coherent states in the system.  In this case either population can reach a homogeneous configuration, $S_\alpha [S_\beta] \sim 1$,
or an inhomogeneous state $(S_\alpha^1 [S_\beta^1] < 0.6)$. Once formed, a localized coherent state is stable. 
These states arise for different partitions of the two populations.

In order to investigate the generality of the phenomenon of localized coherent states, 
we propose another system, denoted as model II,  consisting of $N$ interacting social agents divided
into two populations with sizes $N_\alpha$ and $N_\beta$, such that $N=N_\alpha+N_\beta$.
As in model I, each population constitutes a fully connected network. 
Let the state variable of agent $i \in \alpha [\beta]$ be described by $x_{\alpha[\beta]}(i)$, which can take any of the $q$ values in the set of
natural numbers $\{0, 1,2, . . . , q-1\}$. 
The global fields $M_\alpha$ and $M_\beta$ at a given time are defined as the statistical modes of the states in the populations $\alpha$ and  $\beta$, respectively. These fields can be interpreted as reciprocal opinion trends originated in one population and being transmitted to the other.

The states $x_{\alpha[\beta]}(i)$ in each population are initially assigned at random with a uniform distribution. 
At any given time, a randomly selected agent in population $\alpha[\beta]$ 
can interact either with the global field  $M_{\beta[\alpha]}$ or with any other agent belonging to 
$\alpha[\beta]$, 
in each case according to the dynamics of the bounded confidence model with a threshold value $d$.
We define the dynamics of model II by iterating the following algorithm, similar to that of model I:
\begin{enumerate}
\item Select at random an agent $i\in \alpha$ and a agent $j \in \beta$.
\item Select the source of interaction:
with probability $B$, agent $i \in \alpha$ interacts with field $M_\beta$ and agent $j \in \beta$ interacts with field $M_\alpha$, while
with probability $1-B$,  $i$ interacts with $k \in \alpha$ selected at random and $j$ interacts with $l \in \beta$ also selected at random. 
\item Calculate $|x_{\alpha}(i)-y|$, where $y=M_\beta$ if the source is the field $M_\beta$, or $y=x_\alpha(k)$ if the source is agent 
$k \in \alpha$,  Similarly, calculate $|x_{\beta}(i)-y|$, 
where $y=M_\alpha$ if the source is the field $M_\alpha$, or $y=x_\beta^f(l)$ if the source is agent $l\in \beta$, 
\item If $|x_{\alpha}(i)-y| \leq d$, then set $x_{\alpha}(i)=y$; otherwise $x_\alpha(i)$ does not change. If $|x_{\beta}(i)-y| \leq d$,
then set $x_{\beta}(i)=y$; otherwise $x_\beta(i)$ does not change. 
\item If the source of interaction is $M_{\beta[\alpha]}$, update the fields $M_\alpha$ and $M_\beta$.
\end{enumerate}

\begin{figure}[h]
\begin{center}
\includegraphics[width=0.7\linewidth,angle=0]{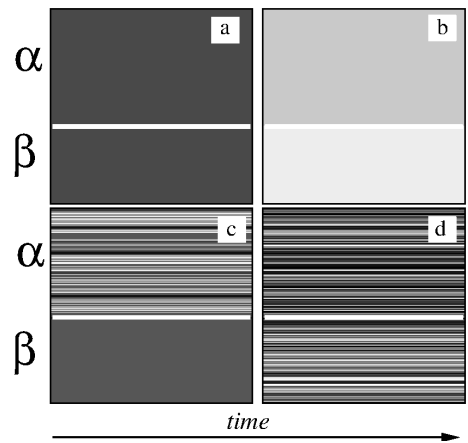}
\end{center}
\caption{Model II. Asymptotic states (vertical axis) of the agents in interacting populations $\alpha$ (upper part) and $\beta$  (lower part) vs. time (horizontal axis) represented in gray colors, for different values of parameters, with fixed $q=200$. 
Each  state variable of an agent is represented by a different shade of gray. Population sizes are $N_\alpha=0.6N$, $N_\beta=0.4N$, with $N=800$.
(a) $d=165, \, B=0.001$ (common homogeneous state). (b) $d=135, \, B=0.00001$ (different homogeneous states).
(c) $d=98, \, B=0.005$ (localized coherent state). (d) $d=25, \, B=0.005$ (disordered state).}
\label{F4}
\end{figure}

Model II displays various collectives behaviors for different values of the parameters $d$ and $B$, similar to those found
in model I. Figure~\ref{F4} shows the asymptotic spatiotemporal patterns associated to the main behaviors observed in model II:
(a) a common homogeneous state, characterized by $M_\alpha=M_\beta$;
(b) different homogeneous states,  with 
 $M_\alpha \neq M_\beta$;
(c) a localized coherent state, where a homogeneous state occurs in only one population while the other is inhomogeneous; 
and (d) a disordered state, where both populations reach inhomogeneous states. 

\begin{figure}[h]
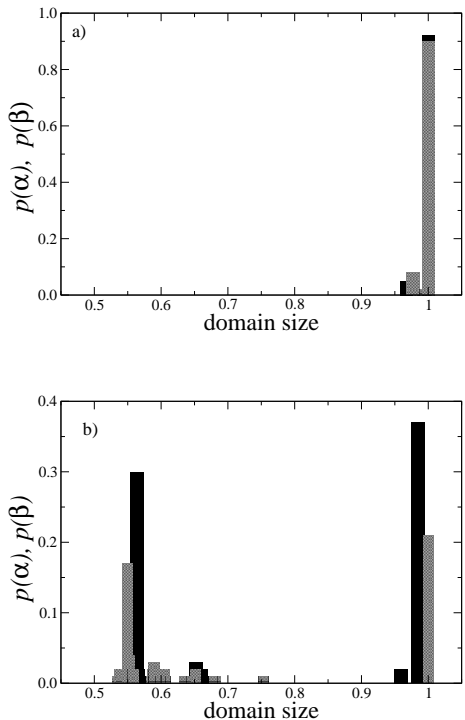

\begin{center}
\includegraphics[width=0.7\linewidth,angle=0]{F5_A.eps}
\end{center}
\vspace{1mm}
\begin{center}
\includegraphics[width=0.7\linewidth,angle=0]{F5_B.eps}
\end{center}
\caption{Model II. Probability distributions  $p(\alpha)$ (black bars) and $p(\beta)$ (gray bars) of normalized domain sizes for populations $\alpha$  and $\beta$, calculated over 
$100$ realizations of initial conditions, with $q=200$, $N_\alpha=0.6N$, $N_\beta=0.4N$, and $N=800$.
(a)  $d= 165, \,  B=0.001$  (common homogenous state); (b) $d=98, \,  B=0.005$ (localized coherent states).}
\label{F5}
\end{figure}

To characterize the localized coherent state in model II, Fig.~\ref{F5} shows the probability distributions $p(\alpha)$ and $p(\beta)$ of the normalized domain sizes for populations $\alpha$ and $\beta$, respectively, for values of $d$ and $B$ corresponding to those of the patterns in 
Figs.~\ref{F4}(a) and \ref{F4}(c). We employ the same criteria as those for Fig.~\ref{F3}. Figure~\ref{F5}(a) 
represents a common homogeneous state characterized by the presence of
one large domain in each population, 
while Fig.~\ref{F5}(b)
reveals the emergence of localized coherent states in the system. In this case, as in model I, either population can reach a homogeneous configuration, or an inhomogeneous state. 

In summary, we have investigated the emergence of localized coherent behavior in 
systems consisting of 
two interacting populations of social agents.
The models that we have considered contain two main ingredients:
(i) the possibility of non-interacting states in the interaction dynamics, 
and (ii) the presence of reciprocal global interactions between the populations.
The global interaction field associated to each population corresponds to the statistical mode of the states of the agents.
In the context of social dynamics, this type of global fields can be interpreted as mass media messages about ``trends'' originated in one population  and being transmitted to the other population. 

In both models, we have found localized coherent states, consisting of one population in a homogeneous state and the other in an disordered state.
These symmetry breaking states arise for different partitions of the two populations. These configurations occur with a probability
that depend on parameters of the system, $B$ and $Q$ in model I, and $B$ and $d$ in model II.
Once it has emerged, a localized coherent state is stable. These states can be considered as intermediate configurations between  a
partially coherent state and a common homogeneous state.

The localized coherent states reported here are reminiscent of the chimera states that have been found in 
two populations of dynamical oscillators having global or long range
interactions, where one population in a coherent state coexist with the other in a incoherent state 
\cite{Abrams,Laing,Scholl,Showalter}. 

In addition, other asymptotic collective behaviors can appear in these systems depending on parameter values:
a common homogeneous state, where both populations share the same state;
different homogeneous states, where both population reach homogenous states but different from each other; 
and a disordered state, where both populations reach inhomogeneous states.

The observation of localized coherent states in the context of social dynamics suggests that the emergence of
chimeralike states should be a common feature in distributed dynamical systems where global interactions coexist with local interactions. 
This phenomenon should also be expected
in other non-equilibrium systems possessing the characteristic of
non-interacting states, such as social and biological systems whose dynamics usually possess a bound
condition for interaction. 
It would also be of interest to search for localized coherent states in complex networks of social agents,
such as communities, where the interaction between populations  occurs through a few elements rather than 
 global fields.  

\section*{Acknowledgments}
J.C.G-A is supported by project No. 500612/2013-7: 151270/2013-9 from CNPq, Brazil. 
M. G. C. is grateful to the Senior Associates Programme of 
the Abdus Salam International Centre for Theoretical Physics, Trieste, Italy, and  acknowledges support from CDCHTA, 
Universidad de Los Andes, Venezuela, under project No. C-1827-13-05-B. 
M. S. M. acknowledges support from Comunitat Aut\`onoma de les Illes Balears, FEDER, and MINECO, Spain, under project No. FIS2007-60327.

\end{document}